\documentclass[twocolumn,showpacs,preprintnumbers,amsmath,
                amssymb,prb,eqsecnum]{revtex4}

\usepackage{graphicx}
\usepackage{dcolumn}
\usepackage{bm}

\begin{document}

\preprint{NpO2/ver.3}

\title{Resonant x-ray scattering spectra from multipole orderings:\\
Np M$_{4,5}$ edges in NpO$_{2}$}

\author{Tatsuya Nagao}
\affiliation{%
Faculty of Engineering, Gunma University, Kiryu, Gunma 376-8515, Japan}

\author{Jun-ichi Igarashi}%
\affiliation{%
Faculty of Science, Ibaraki University, Mito, Ibaraki 310-8512, Japan}

\date{\today}

\begin{abstract}
We study resonant x-ray scattering (RXS) at Np M$_{4,5}$ edges
in the triple-\textbf{k} multipole ordering phase in NpO$_2$, 
on the basis of a localized electron model.
We derive an expression for RXS amplitudes to characterize the spectra
under the assumption that a rotational invariance is preserved in the
intermediate state of scattering process.
This assumption is justified by the fact that energies of the crystal
electric field and the intersite interaction
is smaller than the energy of multiplet structures.
This expression is found useful to calculate energy profiles
with taking account of the intra-Coulomb and spin-orbit interactions.
Assuming the $\Gamma_8$-quartet ground state, we construct 
the triple-\textbf{k} ground state, and analyze the RXS spectra.
The energy profiles are calculated in good agreement with the experiment, 
providing a sound basis to previous phenomenological analyses.

\end{abstract}

\maketitle

\section{\label{sect.1}Introduction}

Resonant x-ray scattering (RXS) technique has attracted much attention
to study spin and orbital properties of 3d transition-metal 
compounds. RXS at the $K$-edge is described by a second-order optical process 
that an incident x-ray excites a $1s$ core electron to unoccupied $4p$ states
and then the $4p$ electron is recombined with the core hole
with emitting x-ray in the dipole process ($E1$).
It became widely known after the observation of
intensities on orbital-ordering superlattice spots at Mn K-edge 
in LaMnO$_3$.\cite{Murakami98.2}
At the earlier stage, the spectra were interpreted 
as a direct observation of orbital ordering.\cite{Ishihara98}
However, subsequent theoretical studies based on band structure 
calculations revealed that the spectra are a direct reflection of 
lattice distortion,\cite{elfimov99,Benfatto99,Takahashi99}
since the $4p$ state in the intermediate state is influenced not 
by the orbital ordering of $3d$
electrons but by lattice distortion through the hybridization with the $2p$
state at neighboring oxygen sites.

Different from transition-metal compounds,
M$_{4,5}$ edges are available for forbidden reflection Bragg spots 
in actinide compounds.\cite{Isaacs90,Mannix99,Longfield02}
The RXS spectra are more directly
reflecting multipole orderings of $5f$ states, since the $E$1 process
involves a transition from the $3d$-core to $5f$ states.
Each actinide atom usually carries local multipole moments, which can order
at low temperatures due to intersite interactions such as exchange 
interactions. For such localized electron systems, RXS amplitudes are given
by summing up contributions at each site.
The crystal electric field (CEF) and the intersite interaction can be 
safely neglected in the intermediate
state, because they are much smaller than the intra-atomic Coulomb interaction. 
Therefore, it may be reasonable to assume that the intermediate state 
preserves the rotational invariance. Under the assumption, 
we derive an expression for the RXS amplitude in the $E$1 process
to characterize the spectra.
Although the expression is essentially the same as the formula
by Hannon {\it et al}.,\cite{Hannon88}
the present form is useful to calculate energy 
profiles with taking full account of multiplet structures. 
Using this expression together with a microscopic model,
we calculate the RXS spectra in the triple-\textbf{k} multipole ordering 
phase in NpO$_2$.

NpO$_2$ undergoes a second-order phase transition below $T_{\textrm 0}=25.5$ K.
\cite{Osborne53,Erdos80}
Since Np ions are Kramers ions in the $(5f)^3$ configuration,
a magnetic ground state is naturally expected.
However, neither M\"{o}ssbauer spectroscopy\cite{Dunlap68,Friedt85} 
nor neutron diffraction experiments\cite{Cox67,Heaton67}
could detect any evidence of the sizable magnetic moment. 
Actually, the former experiment gave an estimate of the upper
limit of the magnitude of the magnetic moment 
$\sim 0.01 \mu_{\textrm B}$, which was too small to explain 
the effective paramagnetic moment $\sim 2.95 \mu_{\textrm B}$.\cite{Ross67}
Another complication is that a muon spin relaxation ($\mu$SR) experiment 
has suggested the low-temperature phase of breaking time-reversal 
symmetry.\cite{Kopmann98} 

A natural way to reconcile with the above 
observations is to introduce the higher-rank multipole ordering rather
than the dipole moment. Actually, Santini and Amoretti proposed a octupole 
ordering of $\Gamma_{2}(xyz)$ symmetry.\cite{Santini00,Santini02}
However, this phase can be ruled out because it gives rise to no RXS 
intensity.
Recently, Paix\~{a}o \textit{et al}. have reported that
a longitudinal triple-\textbf{k} octupole ordering 
accounts well for their RXS experiment.\cite{Paixao02}
The reason for anticipating triple-\textbf{k} orderings is that
it excludes a crystal distortion or a shift of oxygen positions, 
which is consistent with the experiment.
Experimental data obtained from the $^{17}$O NMR spectrum, which 
indicate the existence of two inequivalent oxygen sites,
support the occurrence of the triple-\textbf{k} octupole 
ordering phase.\cite{Tokunaga05}
Some theoretical works also have lent support to
realization of this type of the phase.\cite{Kiss03,Kubo05}

Assuming the $\Gamma_8$-quartet ground state, we explicitly construct 
a triple-\textbf{k} octupole ordering state. This state is found to
simultaneously carry a finite quadrupole moment,
which generates the RXS intensity.
Since the RXS amplitudes are characterized by three terms, the scalar,
dipole, and quadrupole ones, it is not necessary to assume the existence 
of the hexadecapole moment instead of the quadrupole moment.
We calculate the energy profiles with taking full account of multiplet
structures in the intermediate state. 
We obtain spikes-like curves at Np $M_{4}$ edges
for smaller values of the core-level width
$\Gamma$ as a reflection of multiplet structures.
They are found to merge into a single peak with 
$\Gamma >1$ eV.
The energy profile with $\Gamma\sim 2$ eV seems to agree with the experiment.
The azimuthal-angle dependence of the RXS spectra is obtained in agreement
with the previous analysis.\cite{Paixao02,Caciuffo03} 
The present analysis provides a sound basis
to the previous phenomenological analysis.

The present paper is organized as follows.
In Sec. \ref{sect.2}, we present an expression for the RXS amplitude,
which is useful to calculate the energy profiles.
In Sec. \ref{sect.3}, we analyze the RXS spectra in the triple-\textbf{k}
octupole ordering of NpO$_2$ on the basis of a localized electron model.
Section \ref{sect.4} is devoted to concluding remarks.
In Appendix, we derive the general expression of RXS characterizing
energy profiles.

\section{\label{sect.2}Theoretical Framework of RXS}

\subsection{Second-order optical process}

\begin{figure}
\includegraphics[width=6.50cm]{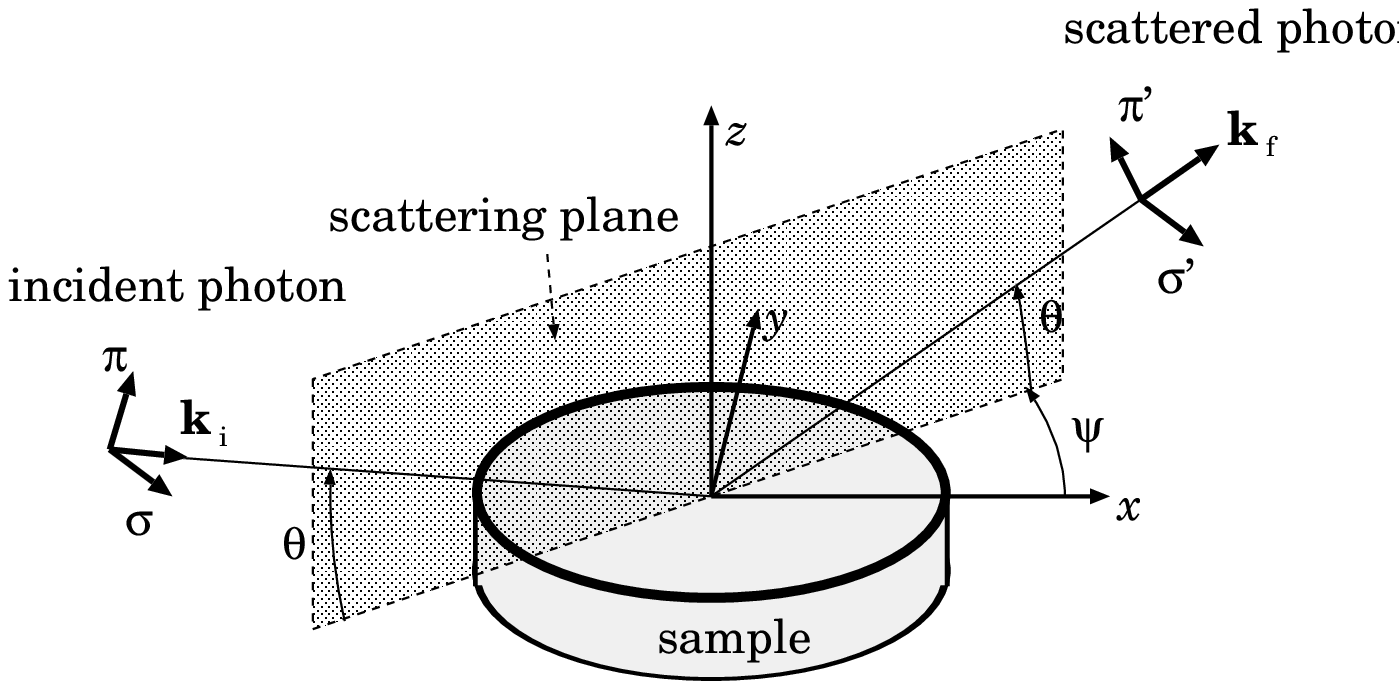}
\caption{\label{fig.geom}
Geometry of the RXS experiment. Photon with polarization 
$\sigma$ or $\pi$ is scattered into the state of polarization
$\sigma'$ or $\pi'$ at the Bragg angle $\theta$.
The azimuthal angle $\psi$ describes the rotation of the sample around 
the scattering vector ${\textbf G}$.
}
\end{figure}

In the resonant process, an incident photon with energy $\hbar \omega$,
wavevector ${\textbf k}$, and polarization vector 
$\mbox{\boldmath{$\epsilon$}}$
excites a core electron to an empty valence shell of the intermediate state, 
then the excited electron falls into the core state emitting a photon 
having the same energy, wavevector ${\textbf k}'$, and polarization vector
$\mbox{\boldmath{$\epsilon$}}'$.
For example, at $M_{4,5}$ edges in actinide compounds, 
a $3d$ core electron is promoted to partially 
filled $5f$ states at each site by the $E$1 transition.
The definition of a geometrical arrangement adopted here is found
in Fig. \ref{fig.geom}.
The RXS amplitude is assumed as a sum of contributions from 
individual ions. 
Since the dipole matrix element involves well-localized wavefunction of core
states, the assumption seems quite reasonable. Accordingly,
the RXS intensity observed in the experiment may be expressed 
for the scattering vector ${\textbf G}$ ($={\textbf k}'-{\textbf k}$) as
\begin{equation}
 I(\mbox{\boldmath{$\epsilon$}}',\mbox{\boldmath{$\epsilon$}}
  ,{\textbf G},\omega) \propto 
   \left|\frac{1}{\sqrt{N}}\sum_{j} {\textrm e}^{-i{\textbf G}\cdot
{\textbf r}_j} 
    M_j(\mbox{\boldmath{$\epsilon$}}',\mbox{\boldmath{$\epsilon$}},\omega)
  \right|^2 ,
\label{eq.rxs.intensity}
\end{equation}
where $M_j(\mbox{\boldmath{$\epsilon$}}',\mbox{\boldmath{$\epsilon$}},\omega)$ 
represents the RXS amplitude at site $j$ with $N$ being the number of sites.
For the $E$1 transition, it is expressed 
as\cite{Blume85,Blume88,Hannon88,Hill96}
\begin{equation}
 M_j(\mbox{\boldmath{$\epsilon$}}',\mbox{\boldmath{$\epsilon$}},\omega) 
 =\sum_{\alpha',\alpha}{\bf \epsilon}'_{\alpha}{\bf \epsilon}_{\alpha'} 
  \sum_{\Lambda} 
  \frac{\langle\psi_0|x_{\alpha, j} |\Lambda\rangle
  \langle \Lambda|x_{\alpha', j}|\psi_0\rangle}
       {\hbar\omega-(E_{\Lambda}-E_0)+i\Gamma}, 
\label{eq.rxs.amplitude}
\end{equation}
where the dipole operators $x_{\alpha, j}$'s are defined as
$x_{1, j}=x_{j}$, $x_{2, j}=y_{j}$, and $x_{3, j}=z_{j}$ 
in the coordinate frame fixed to the crystal axes with the origin located 
at the center of site $j$.
The $| \psi_0 \rangle$ represents the ground state with energy $E_0$, 
while $|\Lambda\rangle$ represents the intermediate state with energy
$E_{\Lambda}$. The $\Gamma$ describes the life-time broadening width 
of the core hole.

\subsection{Energy profiles}

In localized models, the ground state and the intermediate state at each site
are well specified by the eigenfunctions of the angular momentum operator,
$|J,m\rangle$. 
The CEF and the intersite interaction usually lift the degeneracy 
in the ground state. Thus the ground state at site $j$ may be expressed as
$|\psi_0\rangle_j = \sum_m c_j(m)|J,m\rangle$.
On the other hand, in the intermediate state,
we can neglect the CEF and the intersite interaction
in a good approximation, since their energies are much smaller than 
the intra-atomic Coulomb interaction and the spin-orbit interaction (SOI) 
which give rise to the multiplet structure.
Thus the intermediate state preserves the rotational symmetry. 
Under the assumption, as derived in Appendix, 
we obtain a general expression of the scattering amplitude at site $j$:
\begin{eqnarray}
 M_j(\mbox{\boldmath{$\epsilon$}}',\mbox{\boldmath{$\epsilon$}},\omega) 
 &=& \alpha_0(\omega)\mbox{\boldmath{$\epsilon$}}' \cdot
                     \mbox{\boldmath{$\epsilon$}} \nonumber\\
 &-& i\alpha_1(\omega)(\mbox{\boldmath{$\epsilon$}}' 
    \times \mbox{\boldmath{$\epsilon$}})
    \cdot\langle\psi_0|\textbf{J}|\psi_0\rangle \nonumber\\
 &+& \alpha_2(\omega)\sum_\nu 
 P_{\nu}(\mbox{\boldmath{$\epsilon$}}',\mbox{\boldmath{$\epsilon$}})
 \langle\psi_0|z_{\nu}|\psi_0\rangle,
\label{eq.Mtilde.2}
\end{eqnarray}
where
\begin{subequations}
\label{eq.quadrupole1.def}
\begin{eqnarray}
 z_1 &\equiv& Q_{x^2-y^2} = \frac{\sqrt{3}}{2}(J_x^2-J_y^2), \\
 z_2 &\equiv& Q_{3z^2-r^2}= \frac{1}{2}(3J_z^2-J(J+1)), \\
 z_3 &\equiv& Q_{yz}      = \frac{\sqrt{3}}{2}(J_yJ_z+J_zJ_y), \\
 z_4 &\equiv& Q_{zx}      = \frac{\sqrt{3}}{2}(J_zJ_x+J_xJ_z), \\
 z_5 &\equiv& Q_{xy}      = \frac{\sqrt{3}}{2}(J_xJ_y+J_yJ_x), 
\end{eqnarray}
\end{subequations}
and
\begin{subequations}
\label{eq.quadrupole2.def}
\begin{eqnarray}
 P_1(\mbox{\boldmath{$\epsilon$}}',\mbox{\boldmath{$\epsilon$}})
 &=& \frac{\sqrt{3}}{2}(\epsilon'_x\epsilon_x-\epsilon'_y\epsilon_y), \\
 P_2(\mbox{\boldmath{$\epsilon$}}',\mbox{\boldmath{$\epsilon$}})
 &=& \frac{1}{2}(2\epsilon'_z\epsilon_z
                -\epsilon'_x\epsilon_x-\epsilon'_y\epsilon_y), \\
 P_3(\mbox{\boldmath{$\epsilon$}}',\mbox{\boldmath{$\epsilon$}})
 &=& \frac{\sqrt{3}}{2}(\epsilon'_y\epsilon_z+\epsilon'_z\epsilon_y), \\
 P_4(\mbox{\boldmath{$\epsilon$}}',\mbox{\boldmath{$\epsilon$}}) 
 &=& \frac{\sqrt{3}}{2}(\epsilon'_z\epsilon_x+\epsilon'_x\epsilon_z), \\
 P_5(\mbox{\boldmath{$\epsilon$}}',\mbox{\boldmath{$\epsilon$}}) 
 &=& \frac{\sqrt{3}}{2}(\epsilon'_x\epsilon_y+\epsilon'_y\epsilon_x).
\end{eqnarray}
\end{subequations}
Here we have suppressed the dependence on $j$ in the right hand side of
Eq.~(\ref{eq.Mtilde.2}).
The energy profiles are given by only three functions,
$\alpha_0(\omega)$, $\alpha_1(\omega)$, and $\alpha_2(\omega)$,
whose expressions are explicitly given in Appendix.

Several facts are immediately deduced from Eq.~(\ref{eq.Mtilde.2}).
First, since the scalar, dipole, and quadrupole terms exhaust the
amplitude, the octupole ordering alone does not give rise to the RXS
amplitude. Second, the choice of the CEF parameters in the ground state
does not affect the shape of energy profiles $\alpha_0(\omega)$, 
$\alpha_1(\omega)$ and $\alpha_2(\omega)$, 
although it affects the expectation values 
of dipole and/or quadrupole operators.
Third, $\alpha_0(\omega)$ has no contribution to the forbidden Bragg spots
in the antiferro-type structure.
In order to calculate the energy profiles, however, we need to know explicitly
wavefunctions of the intermediate state, which are discussed
in the next section.

\subsection{Absorption coefficient}

Within the $E1$ transition, the absorption coefficient is given by
\begin{equation}
 A(\omega) \propto \sum_j\sum_{\alpha}\sum_{\Lambda}
 |\langle\Lambda|x_{\alpha,j}|\psi_0\rangle|^2
 \frac{\Gamma/\pi}{(\hbar\omega-E_{\Lambda}+E_0)^2+\Gamma^2},
\label{eq.absorption}
\end{equation}
where $|\Lambda\rangle$ with energy $E_{\Lambda}$ represents the final state,
which is equivalent to the intermediate state of RXS.
A comparison of Eq.~(\ref{eq.absorption}) with Eq.~(\ref{eq.rxs.amplitude})
leads to
\begin{equation}
 A(\omega) \propto -{\rm Im}\, \alpha_0(\omega),
\end{equation}
where \textrm{Im} $X$ denotes the imaginary part of $X$.

\section{\label{sect.3}RXS spectra from N\lowercase{p}O$_{2}$} 

\subsection{Quartet ground state}

\begin{figure}
\includegraphics[width=6.50cm]{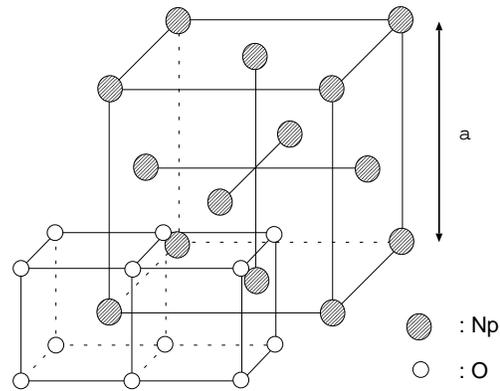}
\caption{\label{fig.struc1} Crystal structure of NpO$_{2}$. 
Gray circles denote Np ions and open circles represent O ions.}
\end{figure}

NpO$_{2}$ has the CaF$_2$ type structure ($F_{m\overline{3}m}$)
with a lattice constant $a=5.431$ \textrm{\AA} at room temperature,
as schematically shown in Fig.~\ref{fig.struc1}.\cite{Osborne53}
Np ions are tetravalent in NpO$_{2}$, as confirmed
by the isomer shift in M\"{o}ssbauer spectra\cite{Dunlap68}
and by the neutron diffraction experiment.\cite{Delapalme80}
In a localized description, each Np ion is in the $(5f)^3$-configuration.
The Hamiltonian of Np ions consists of the intra-atomic Coulomb
interaction between $5f$ electrons in addition to the SOI of $5f$ electrons. 
The Slater integrals for the Coulomb 
interaction and the SOI parameters are evaluated within the Hatree-Fock 
approximation (HFA),\cite{Cowan81} and are listed in Table \ref{table.1}.
Because the isotropic parts of the Coulomb interaction $F^{0}$'s are known
to be well screened in solids compared to those of the anisotropic parts, 
the former quantities are multiplied by a factor 0.25 while 
the latter's are by 0.8.
Within the HFA, the ground state has the ten-fold
degeneracy corresponding to $J=9/2$ multiplet.
The choice of the multiplying factors does not alter this conclusion.
\cite{Nagao03}
Note that these states of $J=9/2$ are slightly deviated from
those of the perfect Russell-Saunders (RS) coupling scheme 
with $L=6$ and $S=3/2$ due to the presence of the strong SOI.
For instance, ${\textbf L}^2$ and ${\textbf S}^2$ take values 39.752 and 3.237
respectively, compared to the RS values 42 and 3.75.

\begin{table}
\caption{\label{table.1}
Slater integrals and spin-orbit interaction parameters
in the $(3d)^{10}(5f)^3$ configuration
within the HF approximation (in units of eV).\cite{Cowan81}}
\begin{ruledtabular}
\begin{tabular}{llll}
 $F^{k}(3d,3d)$    & $F^{k}(3d,5f)$  & $F^{k}(5f,5f)$ & $G^{k}(3d,5f)$  \\
\hline
 $F^0$ \hspace*{0.2cm} 180.1 & 
  &
 $F^0$ \hspace*{0.2cm} 19.61 & 
 \\
 $F^2$ \hspace*{0.2cm} 92.04 & 
 &
 $F^2$ \hspace*{0.2cm} 9.909 &
 \\
 $F^4$ \hspace*{0.2cm} 59.28 &  
 &
 $F^4$ \hspace*{0.2cm} 6.491 &
 \\
                             &
                             &
 $F^6$ \hspace*{0.2cm} 4.769 & 
                             \\  
\hline
$\zeta_{3d}=$ 76.254 & $\zeta_{5f}=$ 0.298  & \\
\end{tabular}
\end{ruledtabular}
\end{table}

In crystal, the ten-fold degeneracy is lifted by the 
CEF. Under the cubic symmetry, the CEF Hamiltonian $H_{\textrm CEF}$ 
may be expressed as
\begin{equation}
H_{\textrm CEF} = B_4 ( O_4^0 + 5 O_4^4) 
                         +B_6 (O_6^0 - 21 O_6^4),
\label{eq.cef}
\end{equation}
where $O_k^q$'s represent Stevens operator equivalence.
Thereby the degenerate levels are split
into one doublet $\Gamma_6$ and two quartets $\Gamma_8^{(1)}$ 
and $\Gamma_{8}^{(2)}$. The level scheme has been analyzed by the inelastic
neutron scattering, which yields an estimate of CEF parameters as
$B_4 = -3.03 \times 10^{-2}$ meV and $B_6 = 2.36 \times 10^{-4}$ meV.
\cite{Amoretti92}
The lowest levels are given by the $\Gamma_8^{(2)}$, 
which is separated about 55 meV from another quartet $\Gamma_{8}^{(1)}$.
Diagonalizing Eq.~(\ref{eq.cef}), we obtain the bases of the lowest
quartet as
\begin{eqnarray}
\left| + \uparrow \right\rangle
& = & c_{1}  \left| + \frac{9}{2} \right\rangle   
                     + c_{2} \left| + \frac{1}{2} \right\rangle   
                     + c_{3} \left| - \frac{7}{2} \right\rangle, 
\label{eq.g81} \\
\left| + \downarrow \right\rangle
& = & c_{1}  \left| - \frac{9}{2} \right\rangle   
                     + c_{2} \left| - \frac{1}{2} \right\rangle   
                     + c_{3} \left| + \frac{7}{2} \right\rangle, 
\label{eq.g82} \\
\left| - \uparrow \right\rangle
& = & c_{4}  \left| + \frac{5}{2} \right\rangle   
                     + c_{5} \left| - \frac{3}{2} \right\rangle,  
\label{eq.g83} \\
\left| - \downarrow \right\rangle
& = & c_{4}  \left| - \frac{5}{2} \right\rangle   
                     + c_{5} \left| + \frac{3}{2} \right\rangle, 
\label{eq.g84}  
\end{eqnarray}
with $c_1=0.2757, c_2=-0.4483, c_3=0.8503, c_4=-0.9751$ and $c_5=0.2216$.
State $|m \rangle$ denotes the eigenstate with $J_z=m$.
Symbols $\tau$ ($=\pm$) and $\sigma$ ($=\uparrow, \downarrow$) are introduced
to represent the state $|\tau,\sigma\rangle$, which distinguish non-Kramers' 
and Kramers' pairs, respectively.

\subsection{Triple-${\textbf k}$ structure}

The four-fold degeneracy in the ground $\Gamma_8^{(2)}$ quartet
may be lifted by the intersite interaction, giving rise to induced 
multipole moments. Actually, several experiments tell us that
the time-reversal symmetry is broken with nearly zero dipole moment
in the ordered phase below $T_0=25.5$ K.\cite{Kopmann98,Tokunaga05}
These observations lead Santini and Amoretti to propose the antiferro ordering
of $T_{xyz}$-type 
($T_{xyz} \equiv \frac{\sqrt{15}}{6}\overline{J_xJ_yJ_z}$).
\cite{Santini00,Santini02}
Here the overline on operators means symmetrization, for example,
$\overline{J_{x}J_{y}^2}=J_x J_y^2 + J_y J_x J_y + J_y^2 J_x$.\cite{Shiina97}.
Unfortunately, this phase would not give rise to the RXS intensities
observed in the experiments.

An important observation is that no external distortion from cubic structure 
exists in the ordered phase, that is, the unit cell remains cubic below $T_0$.
This leads us to consider the triple-\textbf{k} ordering,
since it allows the crystal to keep the cubic symmetry. 
As schematically shown in Fig.~\ref{fig.struc2} (c),
the triple-\textbf{k} structure is defined by all three members
of the star of ${\textbf k}=\langle001 \rangle$
simultaneously present on each site of the lattice;
there are four sublattices 1, 2, 3 and 4 
at $(0,0,0), (\frac{1}{2},\frac{1}{2},0), (0,\frac{1}{2},\frac{1}{2})$ and 
$(\frac{1}{2},0,\frac{1}{2})$, respectively.

\subsubsection{octupole ordering}

We start by the octupole ordering of $\Gamma_{5u}$-type
proposed by Paix\~{a}o \textit{et al}.\cite{Paixao02}
The corresponding octupole operators are defined by
\begin{subequations}
\label{eq.octupole.def}
\begin{eqnarray}
 T_{x}^{\beta}&=&\frac{\sqrt{15}}{6}\overline{J_x(J_y^2-J_z^2)},\\
 T_{y}^{\beta}&=&\frac{\sqrt{15}}{6}\overline{J_y(J_z^2-J_x^2)},\\
 T_{z}^{\beta}  &=& \frac{\sqrt{15}}{6}\overline{J_z(J_x^2-J_y^2)}.
\end{eqnarray}
\end{subequations}

\begin{figure}
\includegraphics[width=6.50cm]{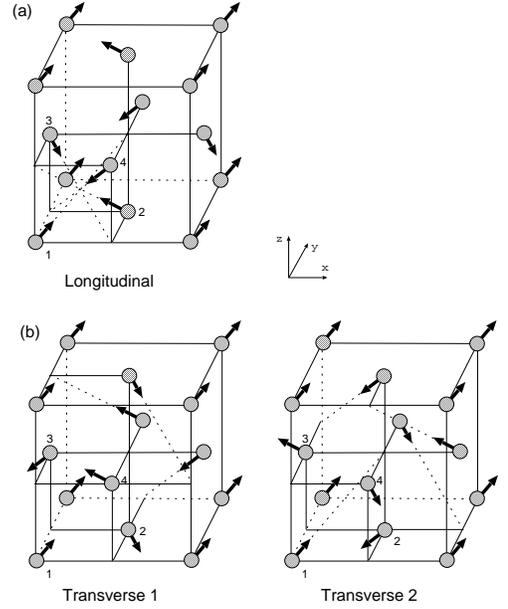}
\caption{\label{fig.struc2} Triple-\textbf{k} antiferro 
arrangements.  (a) a longitudinal pattern and  (b) two transverse patterns
(1 and 2).  Arrows illustrate the direction of the multipole moment
and number 1, 2, 3 and 4 specify the sublattices. 
Oxygen ions are omitted. 
Arrows indicate vector 
$(\langle J_x\rangle,
  \langle J_y\rangle,
  \langle J_z\rangle)$, 
$(\langle Q_{yz}\rangle,
\langle Q_{zx}\rangle,
\langle Q_{xy}\rangle)$, and
$(\langle T^{\beta}_x\rangle,
\langle T^{\beta}_y\rangle,
\langle T^{\beta}_z\rangle)$, 
corresponding to the dipole, quadrupole, and octupole orderings,
respectively.
}
\end{figure}

Defining operators,
\begin{equation}
 T_{p} = \left\{ \begin{array}{lcl}
\frac{1}{\sqrt{3}} \left( T_{x}^{\beta}+T_{y}^{\beta}+T_{z}^{\beta} \right) 
     & {\textrm for} & p=111, \\
\frac{1}{\sqrt{3}} \left( T_{x}^{\beta}-T_{y}^{\beta}-T_{z}^{\beta} \right) 
     & {\textrm for} & p=\overline{1}11, \\
\frac{1}{\sqrt{3}} \left(-T_{x}^{\beta}+T_{y}^{\beta}-T_{z}^{\beta} \right) 
     & {\textrm for} & p=1\overline{1}1, \\
\frac{1}{\sqrt{3}} \left(-T_{x}^{\beta}-T_{y}^{\beta}+T_{z}^{\beta} \right) 
     & {\textrm for} & p=11\overline{1}, \\
\end{array} \right. 
\end{equation}
we assign them to each sublattice.
Each $T_{p}$ operator has eigenvalues $\pm t_1$ ($t_1=-6.102$) 
and doubly degenerated $0$. The eigenstates of eigenvalues $\pm t_1$
are connected to each other by the time-reversal operation,
and so are two degenerate states of eigenvalue 0.
For example, the eigenstate of eigenvalue $-t_1$ for $T_{111}$ is
explicitly written as
\begin{eqnarray}
|-t_1 \rangle &=& 
\frac{1}{2}{\textrm e}^{{\textrm i}\left(\theta_{111}-\frac{\pi}{2}\right)}
       \left[ |+ \uparrow \rangle
 + {\textrm e}^{-{\textrm i}\frac{\pi}{4}} |+ \downarrow \rangle
\right] \nonumber \\
& & 
- \frac{1}{2} \left[ |- \uparrow \rangle
 + {\textrm e}^{ {\textrm i}\frac{3 \pi}{4}} |- \downarrow \rangle
\right],
\end{eqnarray}
with $\theta_{111}$ being an angle between vector $(1,1,1)$ and the $z$ axis,
that is, $\cos \theta_{111} = \sqrt{1/3}$, $\sin \theta_{111} = \sqrt{2/3}$.
This state is different from a ground state assumed by Lovesey {\it et al},
who considered the state deviating from $\Gamma_8$ quartet.\cite{Lovesey03}
Using the eigenstates as bases, $T_p$ is represented as
\begin{equation}
T_p = 
\left(\begin{array}{cccc}
    -t_1 & 0 & 0 & 0 \\
    0 &+t_1 & 0 & 0 \\
    0 & 0 & 0 & 0 \\
    0 & 0 & 0 & 0 \\
\end{array} \right). 
\end{equation}
The ground state is given by assigning either of eigenstates of $\pm t_1$ 
to each sublattice; which eigenstate is relevant depends on the sign of
acting mean field. 
As shown in Fig.~\ref{fig.struc2}, one longitudinal order and two transverse 
orders are possible in the triple-\textbf{k} ordering;
for the longitudinal one, eigenstates of
$T_{111}$, $T_{\overline{1}11}$, $T_{1\overline{1}1}$ and
$T_{11\overline{1}}$ are assigned to sublattices 1, 3, 4 and 2, respectively;
for two transverse orders, eigenstates of of $T_{111}$, $T_{\overline{1}11}$, 
$T_{1\overline{1}1}$, and $T_{11\overline{1}}$, are assigned to
sublattices 1, 2, 3 and 4, and to 1, 4, 2 and 3, respectively.

Introducing the quadrupole operators,
\begin{equation}
 Q_{p} = \left\{ \begin{array}{lcl}
\frac{1}{\sqrt{3}} \left( Q_{yz}+Q_{zx}+Q_{xy} \right) & {\textrm for} & 
     p=111, \\
\frac{1}{\sqrt{3}} \left( Q_{yz}-Q_{zx}-Q_{xy} \right) & {\textrm for} & 
     p=\overline{1}11, \\
\frac{1}{\sqrt{3}} \left(-Q_{yz}+Q_{zx}-Q_{xy} \right) & {\textrm for} & 
     p=1\overline{1}1, \\
\frac{1}{\sqrt{3}} \left(-Q_{yz}-Q_{zx}+Q_{xy} \right) & {\textrm for} & 
     p=11\overline{1}, \\
\end{array} \right. 
\end{equation}
we can construct the quadrupole ordering state by assigning them 
to each sublattice in the same way as for octupole orderings.
Since $Q_p$'s and $T_p$'s are simultaneously diagonalized
because of commuting with each other,
$Q_p$ could be represented as
\begin{equation}
Q_p = 
\left(\begin{array}{cccc}
    -q_1 & 0 & 0 & 0 \\
    0 & -q_1 & 0 & 0 \\
    0 & 0 & +q_1 & 0 \\
    0 & 0 & 0 & +q_1 \\
\end{array} \right),
\end{equation}
with $q_1=-8.273$.

Let the octupole ordering be primarily realized.
Then, each Np ion is in the eigenstate of the eigenvalue $-t_1$ (or $t_1$).
Since the state is also the eigenstate of the eigenvalue $-q_1$,
the quadrupole ordering is simultaneously induced.
On the other hand, if the quadrupole order is primary,
each Np ion is in the eigenstate of the eigenvalue $-q_1$ or $q_1$.
For the case of eigenvalue $-q_1$, two eigensates are to be degenerate and
give eigenvalues $-t_1$ and $t_1$ to the octupole moment $T_p$, 
and thereby the net octupole moment becomes zero. 
For the case of $q_1$, two eigenstates are also to be degenerate
and give the eigenvalue 0 to $T_p$.
In either case, the quadrupole order carries no octupole order.

\subsubsection{dipole ordering}

Although the dipole ordering is ruled out in NpO$_2$, it may be interesting
to discuss here what happens in the dipole ordering.
Introducing the dipole operators,
\begin{equation}
 J_{p} = \left\{ \begin{array}{lcl}
\frac{1}{\sqrt{3}} \left( J_{x}+J_{y}+J_{z} \right) 
     & {\textrm for} & p=111, \\
\frac{1}{\sqrt{3}} \left( J_{x}-J_{y}-J_{z} \right) 
     & {\textrm for} & p=\overline{1}11, \\
\frac{1}{\sqrt{3}} \left(-J_{x}+J_{y}-J_{z} \right) 
     & {\textrm for} & p=1\overline{1}1, \\
\frac{1}{\sqrt{3}} \left(-J_{x}-J_{y}+J_{z} \right) 
     & {\textrm for} & p=11\overline{1}, \\
\end{array} \right.  \label{eq.tripleJ}
\end{equation}
we can construct the dipole ordering state by assigning them to
each sublattice in the same way as in the octupole ordering.
Note that $J_p$ and $Q_p$ are simultaneously 
diagonalized, because both operators commute with each other.
Within the bases of simultaneous eigenstates of $J_p$ and $Q_p$,
the relevant operators are represented as
\begin{eqnarray}
 J_p &=&
     \left(\begin{array}{cccc}
    -j_1 & 0    & 0    & 0    \\
       0 & +j_1 & 0    & 0    \\
       0 & 0    & -j_2 & 0    \\
       0 & 0    & 0    & +j_2 \\
\end{array} \right),  \\
 Q_p &=&
     \left(\begin{array}{cccc}
    -q_{1} & 0      & 0      & 0      \\
         0 & -q_{1} & 0      & 0      \\
         0 & 0      & +q_{1} & 0      \\
         0 & 0      & 0      & +q_{1} \\
\end{array} \right), \label{eq.Qp}  \\
 T_p &=&
     \left(\begin{array}{cccc}
         0 & -t_{1} & 0 & 0 \\
    -t_{1} & 0      & 0 & 0 \\
         0 & 0      & 0 & 0 \\
         0 & 0      & 0 & 0 \\
\end{array} \right),
\label{eq.Tp}
\end{eqnarray}
where $j_1=3.27$, $j_2=0.18$ with parameters given in NpO$_2$.
The magnetic moment is evaluated on either of eigenstates of $\pm j_1$:
$\langle L_p+2S_p\rangle=2.48$ ($L_p$ and $S_p$ are defined as in the same way
as $J_p$). 

In the dipole ordering, the ground state is given by assigning 
one of the eigenstates of $J_p$'s to each sublattice. 
Since $j_1$ is much larger than $j_2$, the ground state is likely to be
either of eigenstates of $\pm j_1$.
It is obvious from Eqs.~(\ref{eq.Qp}) and (\ref{eq.Tp}) 
that the dipole ordering induces the quadrupole moment but no octupole moment.
Note that, if the quadrupole ordering is primary, no dipole moment is induced,
because the doubly-degenerate eigenstates of $Q_p$ are 
the eigenstates of $\pm j_1$ of $J_p$.

\subsection{RXS spectra}

Irrespective of whether the octupole or quadrupole ordering is realized,
RXS amplitudes are generated at each site, 
according to Eq.~(\ref{eq.Mtilde.2}). 
They are proportional 
to $q_1 \alpha_2(\omega)(P_3+P_4+P_5)$ for the simultaneous eigenstate of 
$T_{111}$ and $Q_{111}$,
to $q_1 \alpha_2(\omega)(P_3-P_4-P_5)$ for the simultaneous eigenstate of 
$T_{\overline{1}11}$ and $Q_{\overline{1}11}$,
to $q_1 \alpha_2(\omega)(-P_3+P_4-P_5)$ for the simultaneous eigenstate of 
$T_{1\overline{1}1}$ and $Q_{1\overline{1}1}$,
and to $q_1 \alpha_2(\omega)(-P_3-P_4+P_5)$ for the simultaneous eigenstate of 
$T_{11\overline{1}}$ and $Q_{11\overline{1}}$.
On the scattering vector ${\textbf G}=(h h \ell)$ with $h + \ell=odd$,
these amplitudes are summed up with a positive sign for
sublattices 1 and 2 and with a negative sign for sublattices 3 and 4.
Therefore, the total RXS amplitude becomes proportional to 
$q_1\alpha_2(\omega)P_5$ for the longitudinal order, while they are 
proportional to $q_1\alpha_2(\omega)P_3$ and $q_1\alpha_2(\omega)P_4$
for the two transverse orders.
Note that a similar analysis is applied to the dipole ordering.
In this case, both the dipole and quadrupole terms contribute 
to the amplitude.
These results are summarized in Table \ref{table.6}.
For the transverse case, our present treatment could be extended applying to
the RXS spectra detected at Np $M_4$ edges in 
U$_{0.75}$Np$_{2}$O$_2$.\cite{Wilkins04}
In this compound, the spectra may be interpreted as a consequence
brought about by the transverse type of triple-\textbf{k} AFO ordering
driven by the same ordering pattern at U sites.

\begin{figure}
\includegraphics[width=7.50cm]{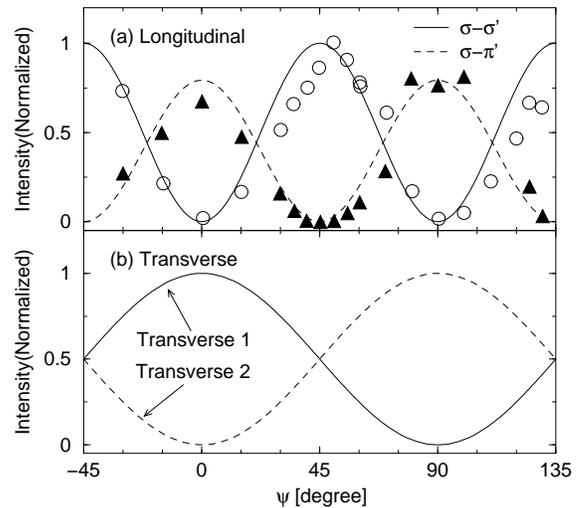}
\caption{\label{fig.azim} Azimuthal-angle dependence of the RXS spectra
from the triple-\textbf{k} AFO phases with ${\textbf G}=(003)$.
(a) Longitudinal ordering. The open circles and filled triangles 
are experimental data, and the solid and broken lines are the calculated 
results, in the $\sigma$-$\sigma'$ and $\sigma$-$\pi'$ 
channels, respectively.\cite{Caciuffo03}
(b) Transverse orderings in the $\sigma$-$\pi'$ channel.
No RXS signal is expected from these orders in the $\sigma$-$\sigma'$ channel.
The solid and broken lines are the calculated results
for the two transverse orders.
}
\end{figure}

Polarization dependences become particularly simple for 
${\textbf G}=(00 \ell)$ ($\ell=$ odd) in the octupole and quadrupole orderings.
They are explicitly written in the scattering geometry shown 
in Fig.~\ref{fig.geom} as 
$P_3=0$, $P_4=0$, $P_5=(\sqrt{3}/2)\sin 2\psi$ in the $\sigma-\sigma'$ channel,
while $P_3=(\sqrt{3}/2)\cos\theta\cos\psi$, 
$P_4=(\sqrt{3}/2)\cos\theta\sin\psi$, 
$P_5=(\sqrt{3}/2)\sin\theta\cos 2\psi$ in the $\sigma-\pi'$ channel.
Figure \ref{fig.azim} shows the azimuthal-angle dependence of the spectra
${\textbf G}=(003)$ in comparison with the experiment.
\cite{Paixao02,Caciuffo03}
The experimental data are well fitted by $\sin^2 2 \psi$ 
in the $\sigma$-$\sigma'$ channel,
and $\sin^2 \theta \cos^2 2 \psi$ in the $\sigma$-$\pi'$ channel.
The two transverse orders cannot reproduce the experimental curves,
as seen from panel (b).
Paix\~{a}o \textit{et al}. and Caciuffo \textit{et al}. analyzed
their experimental data and concluded that the longitudinal order
gives rise to this dependence.\cite{Paixao02,Caciuffo03}
The present analysis confirms their result.
Note that, based on a group theoretical point of view,
Nikolaev and Michel have obtained the same result.\cite{Nikolaev03}

\begin{table*}
\caption{\label{table.6} RXS amplitudes in triple-\textbf{k} ordering,
for ${\bf G}=(hh\ell)$ with $h+\ell=odd$.} 
\begin{ruledtabular}
\begin{tabular}{lccc}
                  &               & RXS amplitude & \\
\hline
\hline
             & Longitudinal  & Transverse 1 & Transverse 2 \\
\hline
dipole  
& -${\rm i}\alpha_1(\omega) \langle \psi_0|J_{z}|\psi_0 \rangle
  (\mbox{\boldmath{$\epsilon$}}' \times \mbox{\boldmath{$\epsilon$}})_z$
& -${\rm i}\alpha_1(\omega) \langle \psi_0|J_{x}|\psi_0 \rangle
  (\mbox{\boldmath{$\epsilon$}}' \times \mbox{\boldmath{$\epsilon$}})_x$
& -${\rm i}\alpha_1(\omega) \langle \psi_0|J_{y}|\psi_0 \rangle
  (\mbox{\boldmath{$\epsilon$}}' \times \mbox{\boldmath{$\epsilon$}})_y$
\\
& $+ \alpha_2(\omega) \langle \psi_0|Q_{5}|\psi_0 \rangle
  P_5 (\mbox{\boldmath{$\epsilon$}}',\mbox{\boldmath{$\epsilon$}})$ 
& $+ \alpha_2(\omega) \langle \psi_0|Q_{3}|\psi_0 \rangle
  P_3 (\mbox{\boldmath{$\epsilon$}}',\mbox{\boldmath{$\epsilon$}})$ 
& $+ \alpha_2(\omega) \langle \psi_0|Q_{4}|\psi_0 \rangle
  P_4 (\mbox{\boldmath{$\epsilon$}}',\mbox{\boldmath{$\epsilon$}})$ \\
\hline
 quadrupole 
& $\alpha_2(\omega) \langle \psi_0|Q_{5}|\psi_0 \rangle
  P_5 (\mbox{\boldmath{$\epsilon$}}',\mbox{\boldmath{$\epsilon$}})$ 
& $\alpha_2(\omega) \langle \psi_0|Q_{3}|\psi_0 \rangle
  P_3 (\mbox{\boldmath{$\epsilon$}}',\mbox{\boldmath{$\epsilon$}})$ 
& $\alpha_2(\omega) \langle \psi_0|Q_{4}|\psi_0 \rangle
  P_4 (\mbox{\boldmath{$\epsilon$}}',\mbox{\boldmath{$\epsilon$}})$ \\
\hline
 octupole 
& $\alpha_2(\omega) \langle \psi_0|Q_{5}|\psi_0 \rangle
  P_5 (\mbox{\boldmath{$\epsilon$}}',\mbox{\boldmath{$\epsilon$}})$ 
& $\alpha_2(\omega) \langle \psi_0|Q_{3}|\psi_0 \rangle
  P_3 (\mbox{\boldmath{$\epsilon$}}',\mbox{\boldmath{$\epsilon$}})$ 
& $\alpha_2(\omega) \langle \psi_0|Q_{4}|\psi_0 \rangle
  P_4 (\mbox{\boldmath{$\epsilon$}}',\mbox{\boldmath{$\epsilon$}})$ \\
\end{tabular}
\end{ruledtabular}
\end{table*}

Now we discuss the energy profiles. In order to calculate them, we need
the wavefunctions in the intermediate state.
We first evaluate the Slater integrals for the Coulomb interaction 
and the SOI parameters within the HFA, which are shown in Table \ref{table.4}.
These values are reduced by taking account of screening effects.
The reduction factors are set the same as in the ground state.
The Hamiltonian of the intermediate state,
consisting of the full intra-atomic Coulomb interactions 
between $5f$-$5f$, $5f$-$3d$ and $3d$-$3d$ electrons as well as the SOI 
of $5f$ and $3d$ electrons, is represented by $1001 \times (2 j_d+1)$ 
microscopic states with the total angular momentum
of the core hole $j_d=3/2$ and $5/2$ corresponding to the
$M_4$ and the $M_5$ edges, respectively.
Diagonalizing the Hamiltonian matrix, we obtain multiplet structures 
in the intermediate state. The $\alpha_2(\omega)$ is calculated by
using Eq.~(A.8).

\begin{table}
\caption{\label{table.4}
Slater integrals and spin-orbit interaction parameters
in the $(3d)^{9}(5f)^4$ configuration
within the HFA (in units of eV).\cite{Cowan81}}
\begin{ruledtabular}
\begin{tabular}{llll}
 $F^0$ \hspace*{0.2cm} 181.0 & 
 $F^0$ \hspace*{0.2cm} 29.13 & 
 $F^0$ \hspace*{0.2cm} 20.54 & 
 $G^1$ \hspace*{0.2cm} 2.158 \\
 $F^2$ \hspace*{0.2cm} 92.62 & 
 $F^2$ \hspace*{0.2cm} 2.749 & 
 $F^2$ \hspace*{0.2cm} 10.39 &
 $G^3$ \hspace*{0.2cm} 1.306 \\
 $F^4$ \hspace*{0.2cm} 59.67 &  
 $F^4$ \hspace*{0.2cm} 1.281 & 
 $F^4$ \hspace*{0.2cm} 6.943 &
 $G^5$ \hspace*{0.2cm} 0.914 \\
                             &
                             &
 $F^6$ \hspace*{0.2cm} 5.017 & 
                             \\  
\hline
$\zeta_{3d}=$ 77.278 & $\zeta_{5f}=$ 0.339  & \\
\end{tabular}
\end{ruledtabular}
\end{table}

\begin{figure}
\includegraphics[width=8.0cm]{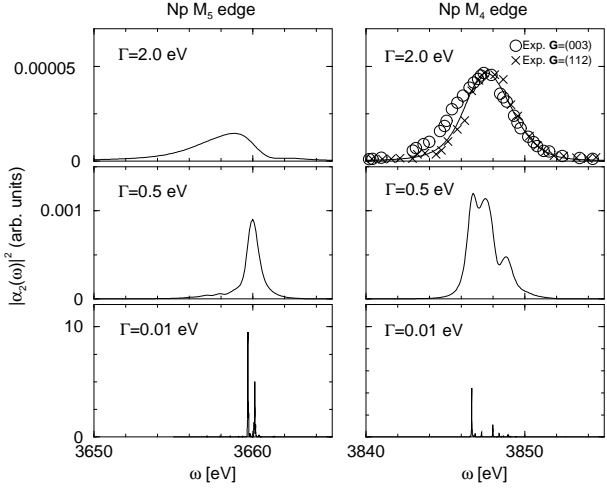}
\caption{\label{fig.alpha_2} 
Energy profiles $|\alpha_2(\omega)|^2$ at the Np $M_4$ 
(right panels) and $M_5$ (left panels) edges. 
The lines represent
the calculated results for $\Gamma=0.01, 0.5$ and $2.0$ eV, respectively,
from bottom to top panels. In the top right panel,
The open circles and crosses are experimental data in 
NpO$_2$.\cite{Paixao02,Caciuffo03} The peak heights of them are
adjusted to that of the calculated value.
}
\end{figure}

The energy profile is proportional to $|\alpha_2(\omega)|^2$
in the octupole ordering phase.
The calculated spectra around $M_4$ and $M_5$ edges are displayed 
with several choices of $\Gamma$ values in Fig. \ref{fig.alpha_2}.
The origin of the energy is adjusted such that the peak of the
RXS spectrum is located at the experimental peak position.
Since there is no reliable estimation for the $\Gamma$ value,
we choose three typical values $\Gamma=0.01, 0.5$ and $2.0$ eV.
The spike-like curves with $\Gamma=0.01$ eV directly reflect the multiplet
splittings of the intermediate states.
For the $M_4$ edge, the choice $\Gamma=0.5$ eV makes a multi-peak-structure
line-shape. It merges into a single-peak structure around 
$\Gamma \simeq 1.0$ eV.
The choice $\Gamma=2.0$ eV corresponds to one of better fittings 
with the experimental line shape.\cite{Paixao02,Caciuffo03} 
The core-level energy is adjusted such that the calculated peak at the $M_4$
edge with $\Gamma=2$ eV coincides with the experimental one.
Paixao {\it et al}. reported that the line shape is well fitted by 
a Lorentzian-squared rather than a Lorentzian one.\cite{Paixao02}
As shown above, the line shape is basically determined by the multiplet
structure, which is smeared by the life-time broadening.
Whether it looks Lorentzian-squared or Lorentzian seems unimportant.
As for the spectra at the $M_{5}$ edge, their shape depends rather
sensitively on the value of $\Gamma$ compared to that at the $M_4$ edge.

\begin{figure}
\includegraphics[width=8.0cm]{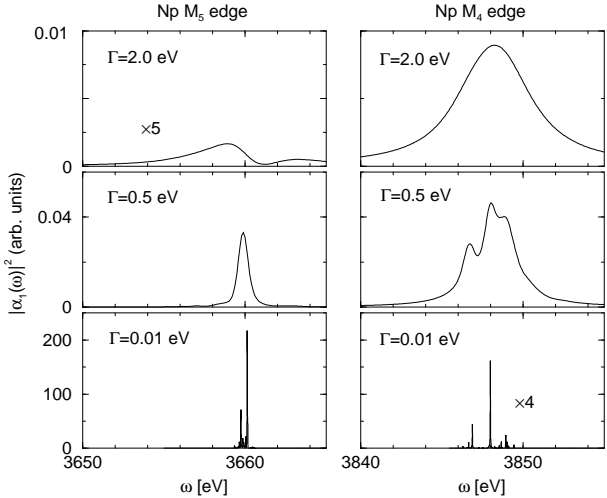}
\caption{\label{fig.alpha_1} 
Energy profiles $|\alpha_1(\omega)|^2$ at the Np $M_4$ 
(right panels) and $M_5$ (left panels) edges. 
The lines represent
the calculated results for $\Gamma=0.01, 0.5$ and $2.0$ eV, respectively,
from bottom to top panels. 
}
\end{figure}

The energy profile in the dipole ordering is given by the sum of the dipole
and quadrupole terms. However, $|\alpha_1(\omega)|^2$ is about two orders of 
magnitude larger than $|\alpha_2(\omega)|^2$.
For instance, $|\alpha_1(\omega)|^2 \sim 192 \times |\alpha_2(\omega)|^2$
when $\Gamma=2.0$ eV.
Thus the dipole term usually dominates the quadrupole term.
Although the dipole ordering is ruled out from experiments,
we show $|\alpha_1(\omega)|^2$ in Fig.~\ref{fig.alpha_1} as a reference.
The peak at the $M_4$ edge with $\Gamma=2$ eV is at 3847.5 eV, 0.7 eV higher
than the peak position of $|\alpha_2(\omega)|^2$.
Note that the spectral shape at the $M_5$ edge depends on $\Gamma$ 
more sensitive than that at the $M_4$ edge.

\subsection{Absorption coefficient}

\begin{figure}
\includegraphics[width=8.0cm]{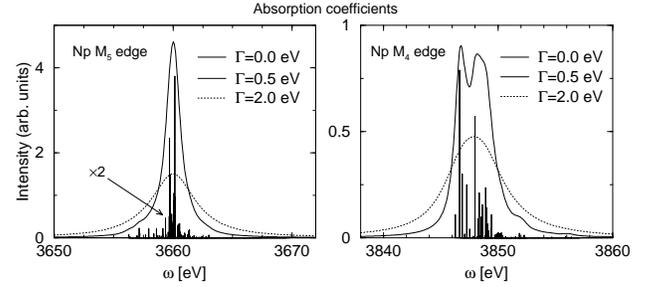}
\caption{\label{Fig.alpha_0} 
Absorption coefficients as functions of the photon energy 
at the Np $M_4$ (right panel) and $M_5$ (left panel) edges. 
The bold solid and dotted lines represent
the calculated results for $\Gamma=0.5$ and $2.0$ eV, respectively.
The vertical bars represent $\delta$-functions with $\Gamma=0$.
}
\end{figure}

The absorption coefficient is proportional to $-{\rm Im} \ \alpha_0(\omega)$.
We calculate $\alpha_0(\omega)$ from Eq.~(A.4) in the same way as
in the calculation of $\alpha_1(\omega)$ and $\alpha_2(\omega)$.
The calculated results are shown in Fig. \ref{Fig.alpha_0} 
at the $M_4$ and $M_5$ edges.
The present calculation confirms the previous multiplet calculation
by Lovesey {\it et al}., in which $-\textrm{Im} \ \alpha_0(\omega)$ has
been calculated at the $M_{\textrm 4}$ 
edge for $\Gamma=0.7$ eV.\cite{Lovesey03}
With increasing values of $\Gamma$, the multiplet structure merges into
a single peak. 
The peak position at the $M_4$ edge with $\Gamma=2$ eV is about $0.35$ eV 
higher than that in $|\alpha_2(\omega)|^2$.

\section{\label{sect.4} Concluding remarks}

In this paper, we have studied the RXS spectra at the Np M$_{4,5}$ edges
in the triple-\textbf{k} multipole ordering phase of NpO$_2$, 
on the basis of a localized electron model. 
We have derived an expression of scattering amplitudes
in the $E$1 process, assuming that
the rotational invariance is preserved in the intermediate states of
the scattering process. This is a reasonable assumption when
the multiplet energy is larger than those of the CEF and the intersite 
interaction.
On the basis of this expression, we have analyzed the RXS spectra in NpO$_2$.
Assuming the $\Gamma_8$-quartet ground state,
we have constructed the triple-\textbf{k} ordering ground state.
The energy profiles have been calculated by taking full account of
the multiplet structure in the intermediate state,
in agreement with the experiment.

RXS signals on multipole ordering superlattice spots have also been observed
and analyzed
at L$_{2,3}$ edges of rare-earth metals in their compounds
such as CeB$_{6}$ and DyB$_{2}$C$_{2}$.\cite{Nakao01,Yakhou01,Tanaka99,
Hirota00,Matsumura05,Nagao01,Igarashi02,Igarashi03}
The intermediate state is created by the transition from the $2p$-core
to $5d$ states.
Since the $5d$ states are considerably delocalized with forming energy bands,
the assumption that the intermediate state preserves the rotational invariance 
becomes less accurate. An extension of the formula is left in future study.

\begin{acknowledgments}
We thank M. Yokoyama and M. Takahashi for valuable discussions.
This work was partially supported by a Grant-in-Aid for Scientific Research 
from the Ministry of Education, Science, Sports and Culture, Japan.
\end{acknowledgments}

\appendix

\section{Derivation of Eq.~(\ref{eq.Mtilde.2})}

We derive a general expression of RXS amplitude 
under the assumption that the intermediate state keeps the rotational 
symmetry at each site. 
The following derivation emphasizes the multiplet
structure in the intermediate state. 
Thereby it is more general than
the previous analyses, in which the fast collision approximation was
adopted by replacing the multiplets with a single level.\cite{Hannon88,Luo93,Lovesey96}
A part of the results found in this Appendix were used
in Ref. \onlinecite{Nagao05} when we analyzed the RXS spectra
form URu$_{2}$Si$_{2}$.

Let the core hole be created at site $j$ in the intermediate state. 
We express the intermediate state as $|\Lambda \rangle=|J',M,i \rangle$,
where the magnitude $J'$ and the magnetic quantum number $M$ 
of total angular momentum (including a core-hole angular momentum)
are good quantum numbers.
To distinguish multiplets having the same $J'$ value but having the 
different energy, we introduce the index $i$.
Defining $M_{\alpha\alpha'}$ by $M_j(\mbox{\boldmath{$\epsilon$}}',
\mbox{\boldmath{$\epsilon$}},\omega)
=\sum_{\alpha\alpha'}\epsilon'_{\alpha}\epsilon_{\alpha'}
M_{\alpha\alpha'}(j,\omega)$,
we rewrite Eq. (\ref{eq.rxs.amplitude}) as
\begin{eqnarray}
 M_{\alpha\alpha'}(j,\omega) 
  &=& \sum_{J',M,i} E_i(\omega,J')
    \langle\psi_0|x_{\alpha, j}|J',M,i \rangle \nonumber \\
    & & \times
          \langle J',M,i|x_{\alpha', j}|\psi_0\rangle,
\label{eq.Mtilde} 
\end{eqnarray}
with
\begin{equation}
E_i(\omega,J') = \frac{1}
       {\hbar\omega-(E_{J',i}-E_0)+i\Gamma}.
\end{equation}
Assuming that the ground-state wavefunction is expressed as 
a linear combination of $| J, m \rangle$ at each site,
\begin{equation}
| \psi_0 \rangle = \sum_{m} c_j(m) | J, m \rangle,
\label{eq.initial}
\end{equation}
and inserting this equation into Eq.~(\ref{eq.Mtilde}),
we obtain
\begin{equation}
 {M}_{\alpha \alpha'}(j,\omega)
 =  \sum_{m,m'} c^{\ast}_j(m) c_j(m') 
   {M}_{\alpha \alpha'}^{m,m'}(\omega), \label{eq.amplitude.2},
\end{equation}
with
\begin{eqnarray}
 {M}_{\alpha \alpha'}^{m,m'}(\omega)
  &=& \sum_{J'} \sum_{i=1}^{N_{J'}} E_i(\omega,J')
    \sum_{M=-J'}^{J'} 
\label{eq.amplitude.3}
\nonumber \\
& & \hspace*{-1.50cm}\times
  \langle J,m|x_{\alpha}|J',M,i \rangle
  \langle J',M,i|x_{\alpha'}| J,m'\rangle,
\end{eqnarray}
where the number of the multiplets having the value $J$ is denoted by $N_J$.
We have suppressed the index $j$ specifying the core-hole site.
The selection rule for the $E$1 process confines the range of the summation
over $J'$ to $J'=J, J \pm 1$. The matrix element of the type
$\langle J, m|x_\alpha|J', M \rangle$ is analyzed by utilizing the
Wigner-Eckart theorem for a vector operator with the use of the 
Wigner's $3j$ symbol;\cite{Tinkham64}
\begin{equation}
 \langle J,m|s_\mu|J'M\rangle = (-1)^{J'+m-1}
\left( \begin{array}{ccc}
J' & 1 & J \\
M & \mu & -m 
\end{array} \right) (J||V_1||J')
\end{equation}
with $s_{\pm 1}=\mp(1/\sqrt{2})(x\pm iy)$, $s_0=z$.
The symbol $(J||V_1||J')$ denotes the
reduced matrix element of the set of irreducible tensor operator of the 
first rank. 
Because of the nature of the dipole operators,
${M}^{m,m '}(\omega) \neq {\textbf 0}$ only when $|m -m '| \leq 2$.
After lengthy calculation, we obtain
\begin{subequations}
\label{eq.tensor}
\begin{eqnarray}
M^{m,m}_{\alpha,\alpha'}(\omega) &=& 
                  \left[\frac{1}{3}J(J+1)-m^2 \right]
                  \alpha_2(\omega) M^{3z^2-r^2}_{\alpha,\alpha'} \nonumber \\
                &-&{\textrm i} m \alpha_1(\omega) M^{z}_{\alpha,\alpha'}
  +\alpha_0(\omega)\delta_{\alpha,\alpha'} , \\
M^{m,m+1}_{\alpha,\alpha'}(\omega) &=& \frac{1}{2}f_m (2m+1)\alpha_2(\omega)
                   (M^{zx}+{\textrm i}M^{yz})_{\alpha,\alpha'} \nonumber \\
                        &-& {\textrm i} \frac{1}{2}f_m \alpha_1(\omega)
                   (M^{x}+{\textrm i}M^{y})_{\alpha,\alpha'},\\
M^{m+1,m}_{\alpha,\alpha'}(\omega) &=& \frac{1}{2}f_m (2m+1)\alpha_2(\omega)
                   (M^{zx}-{\textrm i}M^{yz})_{\alpha,\alpha'} \nonumber \\
                        &-& {\textrm i} \frac{1}{2}f_m \alpha_1(\omega)
                   (M^{x}-{\textrm i}M^{y})_{\alpha,\alpha'},\\
M^{m,m+2}_{\alpha,\alpha'}(\omega) &=& a_m '' \alpha_2(\omega)
                   (M^{x^2-y^2}+{\textrm i}M^{xy})_{\alpha,\alpha'}, \\
M^{m+2,m}_{\alpha,\alpha'}(\omega) &=& a_m '' \alpha_2(\omega)
                   (M^{x^2-y^2}-{\textrm i}M^{xy})_{\alpha,\alpha'},
\end{eqnarray}
\end{subequations}
where
\begin{eqnarray}
f_m &=& \sqrt{(J-m)(J+m+1)}, \\
a_m'' &=& \frac{1}{2} f_m f_{m+1}, 
\end{eqnarray}
and the $3\times 3$ matrices, $M^x$, $M^y$, $M^z$, $M^{xy}$, $M^{yz}$, 
$M^{zx}$, $M^{x^2-y^2}$, and $M^{3z^2-r^2}$ are tabulated 
in Table \ref{table.5}.
\begin{table}
\caption{\label{table.5} 
Antisymmetric matrices, $M^{x}$, $M^{y}$, $M^{z}$, and
symmetric matrices $M^{yz}$, $M^{zx}$, $M^{xy}$,
$M^{3z^2-r^2}$, $M^{x^2-y^2}$.
}
\begin{ruledtabular}
\begin{tabular}{c|ccc}
Antisymmetric           & $M^{x}$ & $M^{y}$& $M^{z}$ \\
& $ \left(\begin{array}{ccc}
 0 & 0 & 0 \\
 0 & 0 & 1 \\
 0 &-1 & 0 \\
\end{array} \right) $
& $ \left(\begin{array}{ccc}
 0 & 0 &-1 \\
 0 & 0 & 0 \\
 1 & 0 & 0 \\
\end{array} \right) $
& $\left(\begin{array}{ccc}
 0 & 1 & 0 \\
-1 & 0 & 0 \\
 0 & 0 & 0 \\
\end{array} \right)$ \\
\hline
Symmetric           & $M^{yz}$ & $M^{zx}$& $M^{xy}$ \\
& $ \left(\begin{array}{ccc}
 0 & 0 & 0 \\
 0 & 0 & 1 \\
 0 & 1 & 0 \\
\end{array} \right) $
& $ \left(\begin{array}{ccc}
 0 & 0 & 1 \\
 0 & 0 & 0 \\
 1 & 0 & 0 \\
\end{array} \right) $
& $\left(\begin{array}{ccc}
 0 & 1 & 0 \\
 1 & 0 & 0 \\
 0 & 0 & 0 \\
\end{array} \right)$ \\
 & & & \\
           & $M^{3z^2-r^2}$ & $M^{x^2-y^2}$&  \\
& $ \left(\begin{array}{ccc}
-1 & 0 & 0 \\
 0 &-1 & 0 \\
 0 & 0 & 2 \\
\end{array} \right) $
& $ \left(\begin{array}{ccc}
 1 & 0 & 0 \\
 0 &-1 & 0 \\
 0 & 0 & 0 \\
\end{array} \right) $
& \\
\end{tabular}
\end{ruledtabular}
\end{table}
The energy profiles are given by
\begin{subequations}
\begin{eqnarray}
\alpha_0(\omega)&=& \frac{2}{3}J(2J-1)F_{J-1}(\omega)
+ \frac{2}{3}J(J+1)F_{J}(\omega) \nonumber \\
  &+& \frac{2}{3}(2 J^2+5J+3)F_{J+1}(\omega),  \\
\alpha_1(\omega)&=& -(2J-1)F_{J-1}(\omega)
- F_{J}(\omega) \nonumber \\
  &+& (2J+3)F_{J+1}(\omega),  \\
\alpha_2(\omega)&=& \frac{4}{3} \left[ -F_{J-1}(\omega)+ F_{J}(\omega)
  - F_{J+1}(\omega) \right], 
\end{eqnarray}
\end{subequations}
with
\begin{equation}
F_{J'}(\omega)= (-)^{J-J'}|(J||V_1||J')|^2
 \sum_{i=1}^{N_{J'}} E_i(\omega,J').
\end{equation}
Substituting Eqs. (\ref{eq.tensor}) into Eq. (\ref{eq.amplitude.2}),
we obtain the final expression Eq.~(\ref{eq.Mtilde.2}).

\bibliography{paper.npo2}

\end{document}